\documentclass[sigconf]{acmart}
\usepackage{graphicx}
\usepackage{amsmath}
\usepackage{amsthm}

\usepackage{amssymb}
\usepackage{amsfonts}
\usepackage{bm}
\usepackage{caption}
\usepackage{enumitem}
\usepackage{booktabs}
\usepackage{graphicx}
\usepackage[textfont={md},labelfont={md}]{subfig}
\usepackage{multirow}

\usepackage[utf8x]{inputenc} 

\renewcommand\footnotetextcopyrightpermission[1]{}
\AtBeginDocument{%
  \providecommand\BibTeX{{%
    \normalfont B\kern-0.5em{\scshape i\kern-0.25em b}\kern-0.8em\TeX}}}

\begin{document}

\fancyhead{}

\title{HQANN: Efficient and Robust Similarity Search for Hybrid Queries with Structured and Unstructured Constraints}

\author{Wei Wu}
\authornote{These authors contributed equally to this work.}
\affiliation{%
  \institution{Heterogeneous Computing Center}
  \city{Kuaishou Technology}
  \country{China}
}
\email{wuwei10@kuaishou.com}

\author{Junlin He}
\authornotemark[1]
\affiliation{%
  \institution{Multimedia Understanding}
  \city{Kuaishou Technology}
  \country{China}
}
\email{hejunlin03@kuaishou.com}

\author{Yu Qiao}
\affiliation{%
  \institution{Multimedia Understanding}
  \city{Kuaishou Technology}
  \country{China}
}
\email{qiaoyu@kuaishou.com}

\author{Guoheng Fu}
\affiliation{%
  \institution{Multimedia Understanding}
  \city{Kuaishou Technology}
  \country{China}
}
\email{fuguoheng@kuaishou.com}

\author{Li Liu}
\affiliation{%
  \institution{Heterogeneous Computing Center}
  \city{Kuaishou Technology}
\country{China}
}
\email{liliu@kwai.com}

\author{Jin Yu}
\authornote{Corresponding author.}
\affiliation{%
  \institution{Multimedia Understanding} 
  \city{Kuaishou Technology}
    \country{China}
}
\email{yujin07@kuaishou.com}

\begin{abstract}
  The in-memory approximate nearest neighbor search (ANNS) algorithms have achieved great success for fast high-recall query processing, but are extremely inefficient when handling hybrid queries with unstructured ($i.e.$, feature vectors) and structured ($i.e.$, related attributes) constraints. In this paper, we present HQANN, a simple yet highly efficient hybrid query processing framework which can be easily embedded into existing proximity graph-based ANNS algorithms. We guarantee both low latency and high recall by leveraging navigation sense among attributes and fusing vector similarity search with attribute filtering. Experimental results on both public and in-house datasets demonstrate that HQANN is 10x faster than the state-of-the-art hybrid ANNS solutions to reach the same recall quality and its performance is hardly affected by the complexity of attributes. It can reach 99\% recall@10 in just around 50 microseconds On GLOVE-1.2M with thousands of attribute constraints.
\end{abstract}



\maketitle

\section{Introduction}
Nearest neighbor search for unstructured data such as texts, images, and videos has played an important role in large scale information retrieval systems.~\cite{s1, s2, s3, s4} Typically, Unstructured data is first converted into high dimensional feature vectors, and subsequent similarity search is conducted on these vectors. As the cost of exact solutions for K-nearest neighbor search~\cite{s5} is extremely high in big data scenarios, researchers have proposed many kinds of approximate nearest neighbor search (ANNS) algorithms~\cite{OPQ10, OPQ6, IVFADC, IVFADCGP, diskann, FAISS, scann}. However, most of the algorithms can only treat hybrid queries with structured and unstructured constraints as disjoint tasks, leading to high query latency and low recall rate, which contradicts the growing demand for hybrid query processing in modern recommendation systems~\cite{vearch, adbv, s6}. Hybrid queries not only supplements the necessary condition information for retrieval, but also effectively improves the accuracy of ANNS algorithms for narrow down the vector search space.

\begin{figure}[t]
\centering
\includegraphics[width=0.47\textwidth,height=3.6cm]{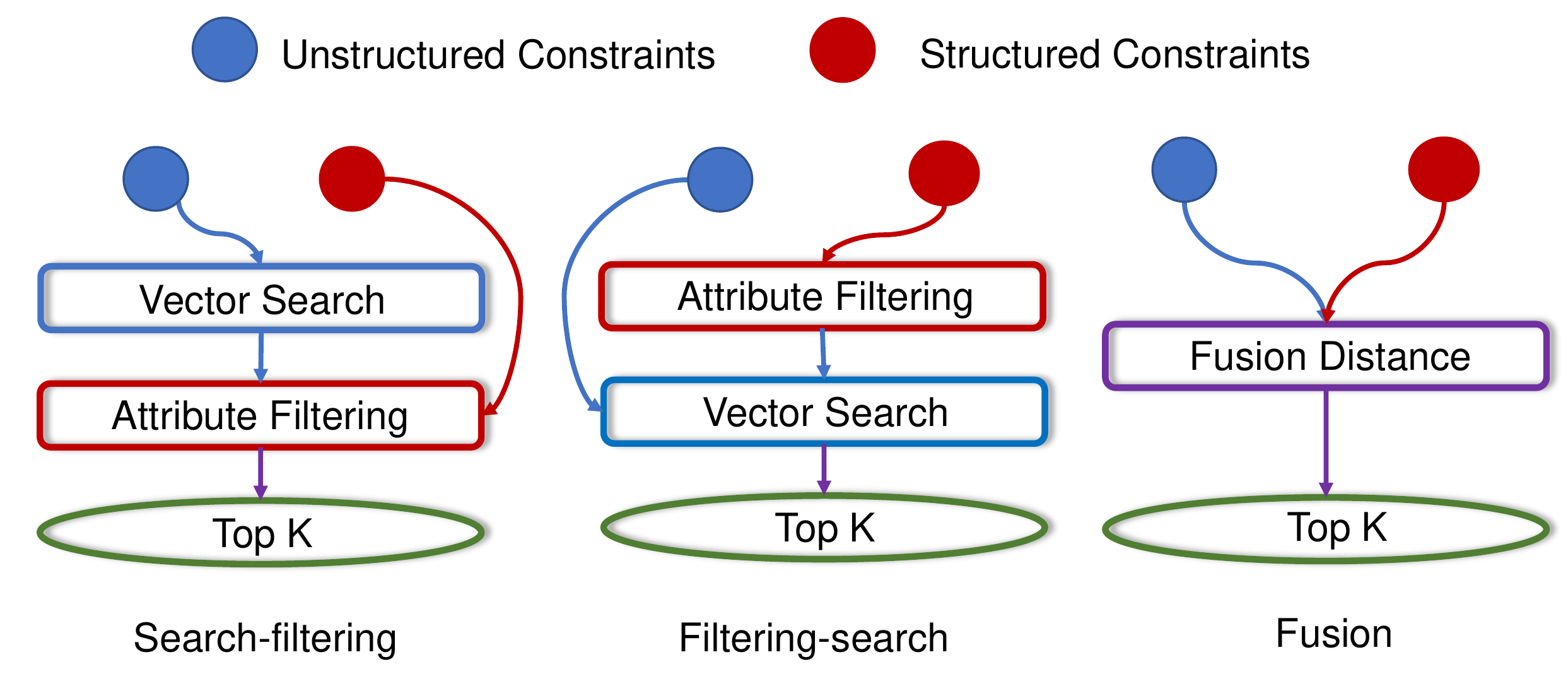}
\caption{Three strategies for hybrid query processing.}
\label{f1}
\end{figure}

There are only a few approaches working on hybrid query processing, including Vearch\cite{vearch}, Alibaba AnalyticDB-V (ADBV)~\cite{adbv} and  Milvus\cite{milvus}. Vearch first conducts the original vector similarity search over the feature vectors and then filters these results with attributes, which can be easily extended to any ANNS algorithms. ADBV introduces a product quantization (PQ)~\cite{OPQ10} based hybrid query framework and enables vector similarity search with a pre-defined bitmap from attribute filtering. Milvus adds an additional partition stage based on frequently used attributes and split the dataset into several subsets, leading to a search space shrinking with attribute constraints. Both of them treat vector similarity search and attribute filtering as separate problems, as shown in Figure ~\ref{f1}, which is not optimal in terms of efficiency or accuracy. The search-filtering strategy has to expand the list of candidates to compensate for the merging loss from two separate subquery systems, while the filtering-search strategy could suffer high latency for exhaustive scanning with a bitmap. Furthermore, the current state-of-the-art PQ-based hybrid query processing systems fail to 
take advantages of proximity graph-based ANNS, which have been proved far superior to PQ in speed and precision without memory concern.

Recently, ~\cite{nhq} proposed a native hybrid query(NHQ) framework based on proximity graph. In NHQ, attribute values processed with $xor$ functions are used as fine-tuning factors for the original distance metric, providing a specialized composite index and joint pruning modules for hybrid queries. However, NHQ regards the original vector distance metric as dominant and fails to emphasize the importance of attributes, its performance drops dramatically as the number of attributes increases.

In this paper, we argue that attributes should be in the leading position for proximity graph-based hybrid query processing. We propose HQANN, a simple yet efficient hybrid query processing framework which can be easily embedded into any existing proximity graph-based ANNS algorithms. Following the fusion strategy in Figure ~\ref{f1}, we merge the search and filtering stage by going through the composite graph, leading to high recall hybrid ANNS algorithms with negligible extra computation overhead.

The main contributions of this paper are as follows:
\begin{itemize}[leftmargin=*]
    \item [$\bullet$] We introduce a novel hybrid query processing framework, HQANN, which efficiently manages hybrid queries in a fused way. HQANN is generally applicable to many proximity graph-based ANNS algorithms. We demonstrate the flow of construction and search for proximity graphs can be efficiently adapted to the proposed framework.
    \item [$\bullet$] We propose to leverage the navigation sense among attributes to traverse the proximity graphs. We present the effectiveness of our method through analysis and experiments.
    \item [$\bullet$] We show that HQANN leads to large performance gains over previous hybrid query processing techniques. Our method achieves state-of-the-art performance on standard large-scale benchmarks such as Glove-1.2M. In additional to recall gains, HQANN greatly improves retrieval efficiency of original proximity graphs, with attribute constraints limiting the search space.
\end{itemize}

\section{Related works}
\subsection{Content-based Retrieval Systems}
Content-based Retrieval systems have made great achievements in facilitating
analytics on unstructured data, such as words, images and videos. In these systems, the learned embedding functions are used to map items in a common vector space, and subsequent retrievals are conducted.~\cite{s7, s8} Although content-based retrieval system effectively support unstructured data, there are many scenarios where both unstructured and structured data should be jointly retrieved(we call them hybrid queries), a typical application is search under a specific limit($i.e$ male or female ). Besides, imposing structured constraints can narrow down the original vector search space and effectively improve the accuracy. 

\subsection{Methods for optimizing hybrid query processing}
Based on the previously proposed high quality ANNS algorithms, Vearch~\cite{vearch}, Alibaba AnalyticDB-V (ADBV)~\cite{adbv} and  Milvus~\cite{milvus} deployed two-stage hybrid processing systems, which treat queries over feature vectors and attributes as disjoint tasks. These systems are only friendly to space partitioning~\cite{IVFADC, IVFADCGP, IMI} or quantization~\cite{opq, lopq, aq} based ANNS solutions, for they have to conduct a detailed scan of the entire dataset with the blacklist from attribute filtering. To take advantage of proximity graph-based ANNS solutions~\cite{HNSW, knn}, NHQ~\cite{nhq} is proposed to provide a specialized composite index and joint pruning modules for hybrid queries. 

Our work differs from the above methods as they fail to insert filtering stage naturally into search stage, while we develop an approach in the following section where we conduct a joint search strategy in a proximity graph.

\section{HQANN}
The performance of hybrid query processing is largely determined by the efficiency of settling attributes, as they determine the accurate search space for feature vectors. From this motivation, we propose HQANN. In HQANN, all attributes is treated as a native part of the fusion distance metric, which can make full use of the inherent Navigation feature of proximity graph-based ANNS solutions.

\subsection{Fusion Distance Metric}
The key point for hybrid query processing is to define an effective distance metric to  facilitate analytics on attributes. Give the fused distance function as the basic premise, a composite proximity graph-based index can be constructed and retrieved without invading its original execution flow.

Generally, performance of a proximity graph is largely determined by the quality of the graph node neighborhoods. Since the feature vector search space of hybrid queries is determined by the attributes, we treat attributes as the dominant part of the fusion distance metric, which means a graph node tends to establish links with datapoints containing the same or vary similar attributes. In such circumstances, a hybrid query can quickly locate those graph nodes with exactly the same attributes and conduct detailed vector similarity search among their neighborhoods. 

Given an hybrid dataset $S$, with m-dimensional feature vector data space $X$ and n-dimensional attribute data space $V$ (we assume that the attributes have been pre-processed into vectors), The distance between two hybrid datapoints $s_i, s_j \in S$ can be measured as: 
\begin{equation}
    Dist(s_i, s_j) = Dist(x_i, x_j) + Dist(v_i, v_j).
\end{equation}
Here $Dist(x_i, x_j)$ can represent any frequently-used vector distance metric($i.e.$ euclidean distance). Let $g(x, y)$ and $f(x, y)$ denote distance functions for feature vectors and attributes separately, and the fusion distance metric can be computed with:
\begin{equation}
    Dist(s_i, s_j) = w \cdot g(x_i, x_j) + f(v_i, v_j),
\end{equation}
where $f(v_i, v_j)$ satisfies
\begin{equation}
    f(v_i, v_j) = 
    \begin{cases}
        0,&  v_i = v_j \\
        bias - \frac{1}{\lg(e(v_i,v_j) + 1)},&  v_i \ne v_j
    \end{cases}
    ,
\end{equation}
with $e(v_i, v_j)$ and $bias$ defined as follows:

\begin{equation}
\begin{split}
    & e(v_i, v_j) = \sum_{k=1}^{n} |v_i[k] - v_j[k]| \\
    & bias \gg \max(w \cdot g(x_i, x_j)) + \frac{1}{lg2}
\end{split}
\end{equation}

Note that attribute vectors should only contain integers in production environments, thus $\min(e(v_i, v_j) = 1)$. We see that when two hybrid datapoints share the same attributes, their attribute distance is defined as zero. Otherwise, $bias$ is applied first to keep attribute distance much larger than feature vector distance, which guarantees the closet neighbors of a hybrid datapoint should have exactly matched attributes. Then $\frac{1}{lg(e(v_i, v_j)+1)}$ is attached as a fine-tuning effect for attribute distance values, leading to smaller distances among attributes with higher similarity. After we obtain the distances among attribute parts, they are fused with feature vector distances with a scale factor $w$,  we tune $w$ to ensure the feature vector distances are much smaller than attribute distances, while avoid them being too small to be ignored.

The mapping function $e(x)$ we used here represents the Manhattan Distance among two attribute vectors, as we only focus on the representation space of attribute distances instead of their accurate values. For example, if $e(v_i, v_j)$ preforms a $xor$ operation for all dimension of two attribute vectors and sum them up, then return value of $f(v_i, v_j)$ would be the same for a mass of combinations of attributes, this deprives the navigation sense among attributes and the composite proximity graph will degenerate into a normal one with feature vector similarity information only. In fact, any mapping function that is capable of preserving attribute representation space can be used as $e(x)$ ($i.e.$ Euclidean distance), we select Manhattan Distance here for computational efficiency. The $lg$ function is applied to prevent attribute values from being too large and erases the gap among different attributes.

Eq.2-4 provide a simple and efficient solution to fuse two disjoint distance metrics. In our experimental study, $w = 0.25$ and $bias = \max(g(x_i, x_j)) + \frac{1}{\lg2} $ work well for most datasets. We want to point out that in most production scenarios these feature vectors $x \in X$ are pre-normalized with inner product(IP) as the distance metric, which means $\max(g(x_i, x_j)) = 1$, thus we do do not need to tune these hyper parameters specifically for most cases (to satisfy minimum priority in fusion distance metric, let $g(x_i, x_j) = 1 - x_i \cdot x_j$ for IP).

\subsection{Composite Proximity Graph}
The basic idea of HQANN is that we obtain a fusion distance metric by fine-tuning the attribute distances via feature vector distances. More specifically, the distance between two hybrid datapoints is mainly defined by their attribute distance, and their feature vector distance serves as a adjustment factor with limited impact on the fusion distance. During the construction stage of the proximity graphs, datapoints with the same or very similar attributes will be linked first as they tend to have much smaller distance compared with our candidates. Then the remaining neighborhood vacancies would be filled up with datapoints that are relatively distant in attributes, which strongly maintains the connectivity of the graph. In the search stage, a hybrid query traverses the graph until it reaches the first node with the same attribute, by fully scanning the node's neighborhoods, a high quality return candidate list can be efficiently accessed.

\begin{figure}[t]
  \includegraphics[width=0.47\textwidth, height=0.3\textwidth]{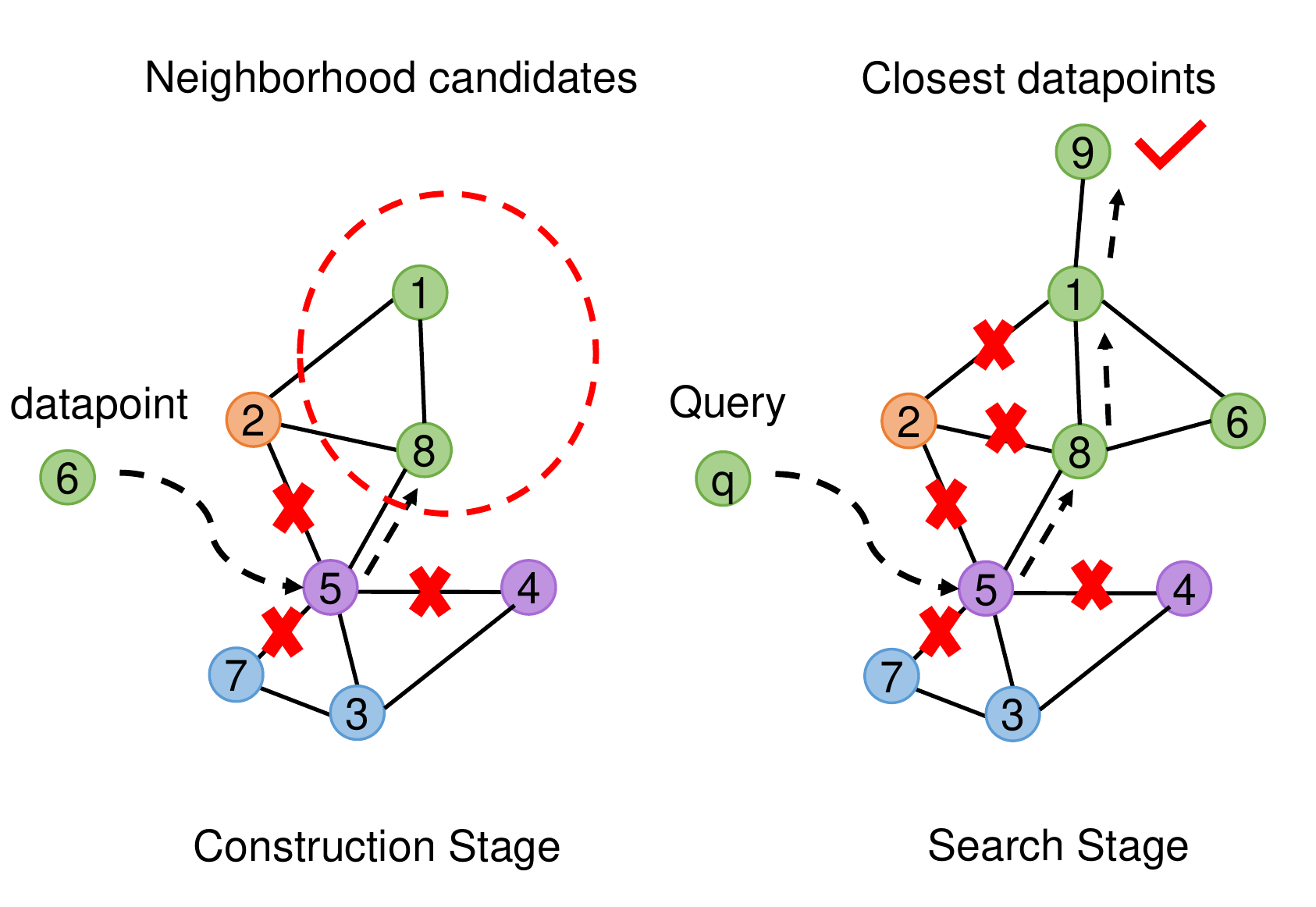}
  \caption{Work flow of HQANN framework.}
  \label{f2}
\end{figure}

The overall configuration of the composite proximity graph is shown in Figure ~\ref{f2}. With attribute distances play a major role in the fusion distance metric, we actually accomplish the filtering-search strategy within a single traversing of the graph. The numerical differences among attribute distances and feature vector distances skillfully exploiting the proximity graphs' essence of voracious searching for graph nodes closest to the query. Hence, HQANN is robust against scale of attributes and remains highly efficient.

\section{Experiments}
In this section, we demonstrate that HQANN is more efficient and robust compared with previous state-of-the-art hybrid processing methods, and show that under attribute constraints, the composite graph surpasses the original proximity graph in both latency and recall rate.

\subsection{Experiment Setup}
We conduct all the experiments on an Intel Xeon E5-2680(2400MHz frequency) with 16 CPU threads. The datasets we used are four public datasets GLOVE-1.2M, SIFT-1M, GIST-1M, DEEP-1B from the ANNS benchmark website~\cite{annbenchmark}, and an extra billion-scale in-house merchandise dataset with millions of merchant IDs as attribute constraints. As public datasets do not originally contain structured constraints, we generate attributes randomly for all datapoints following the same method in~\cite{milvus}. All datapoints are pre-normalized and we set $w = 0.25, bias = 4.32$ by default unless otherwise specified. 

\begin{figure*}[ht]
\centering
\subfloat[Recall@10 (GLOVE-1.2M)]{
\includegraphics[width=4.2cm,height=2.8cm]{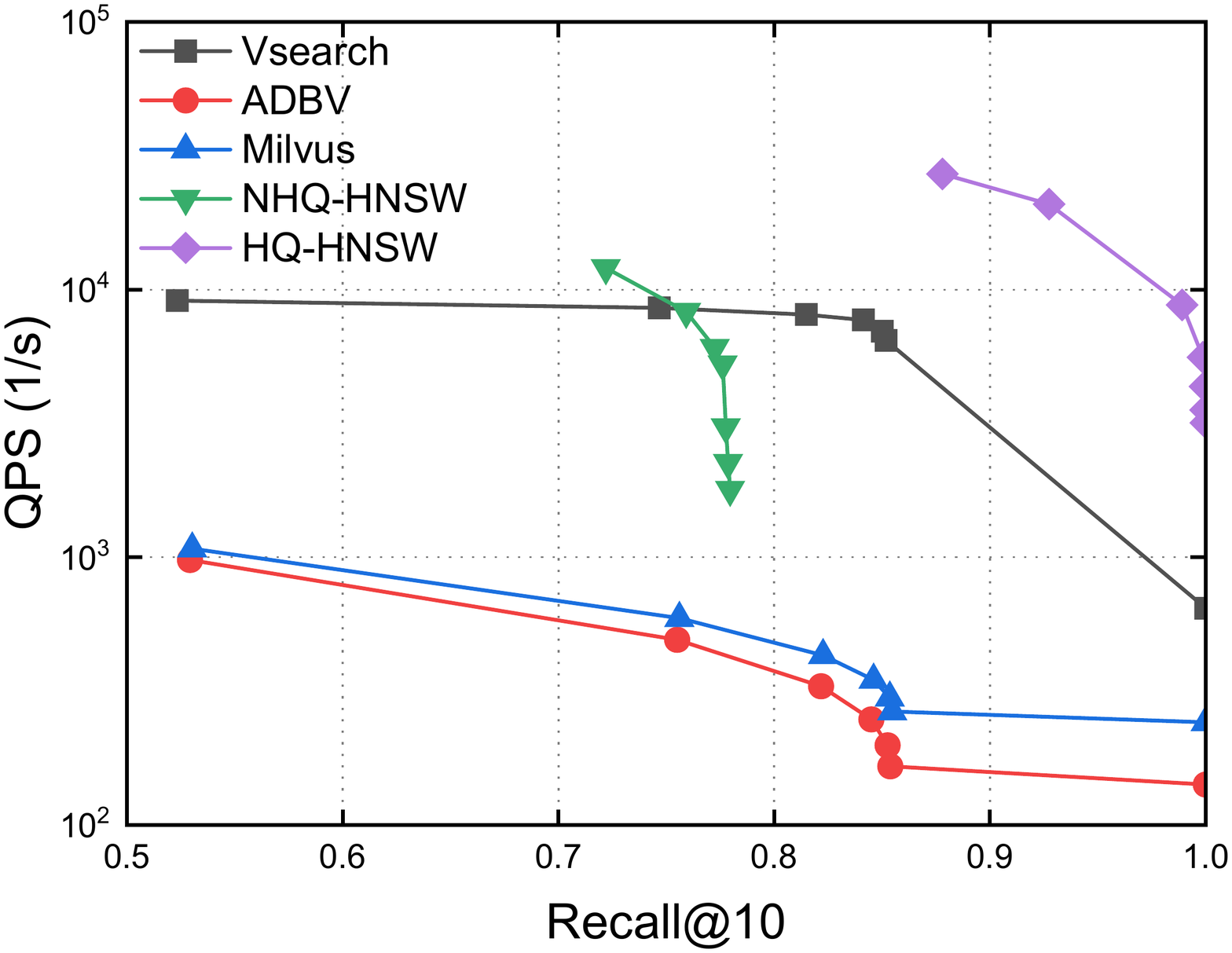}
}
\subfloat[Recall@10 (SIFT-1M)]{
\includegraphics[width=4.2cm,height=2.8cm]{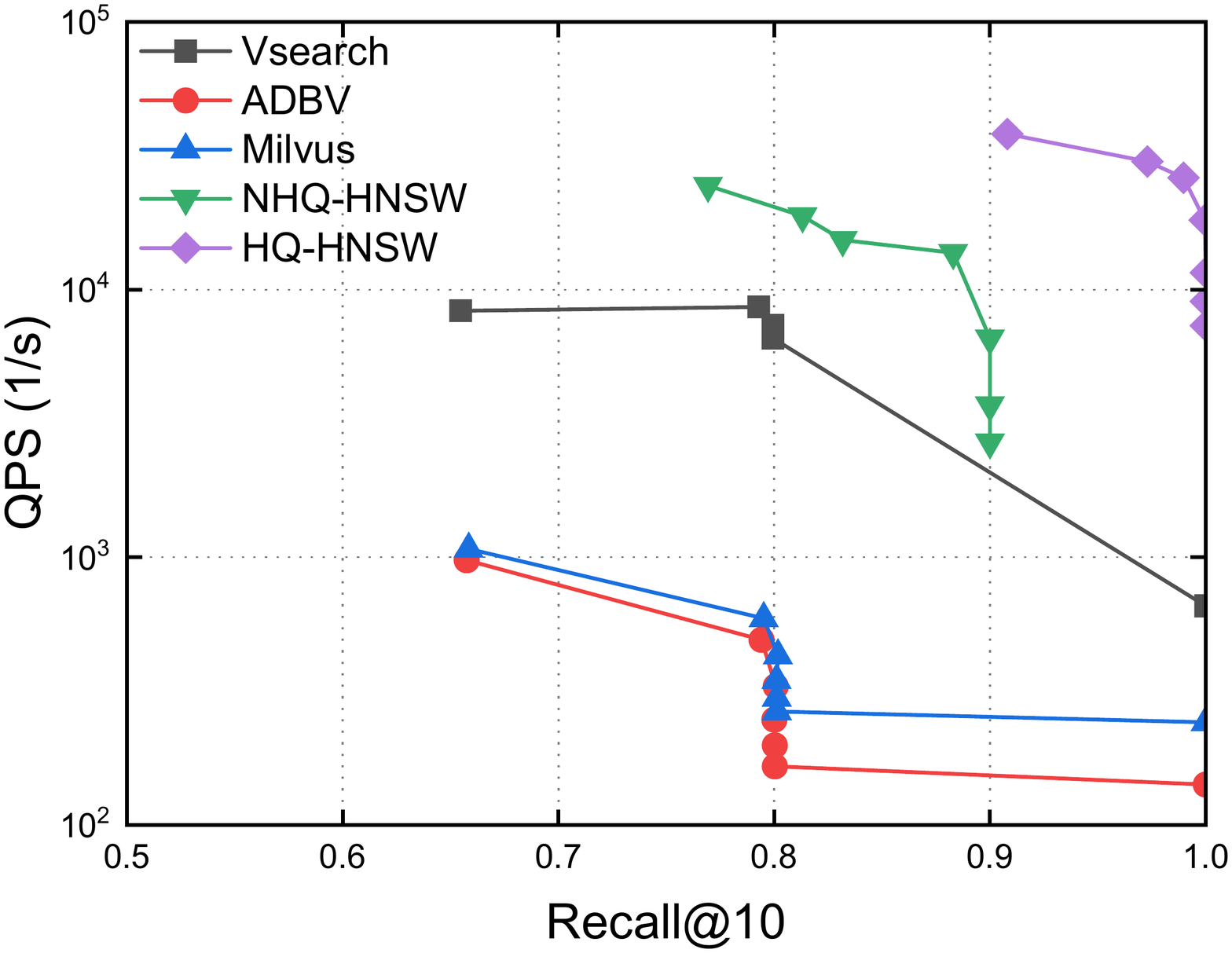}
}
\subfloat[Recall@10 (GIST-1M)]{
\includegraphics[width=4.2cm,height=2.8cm]{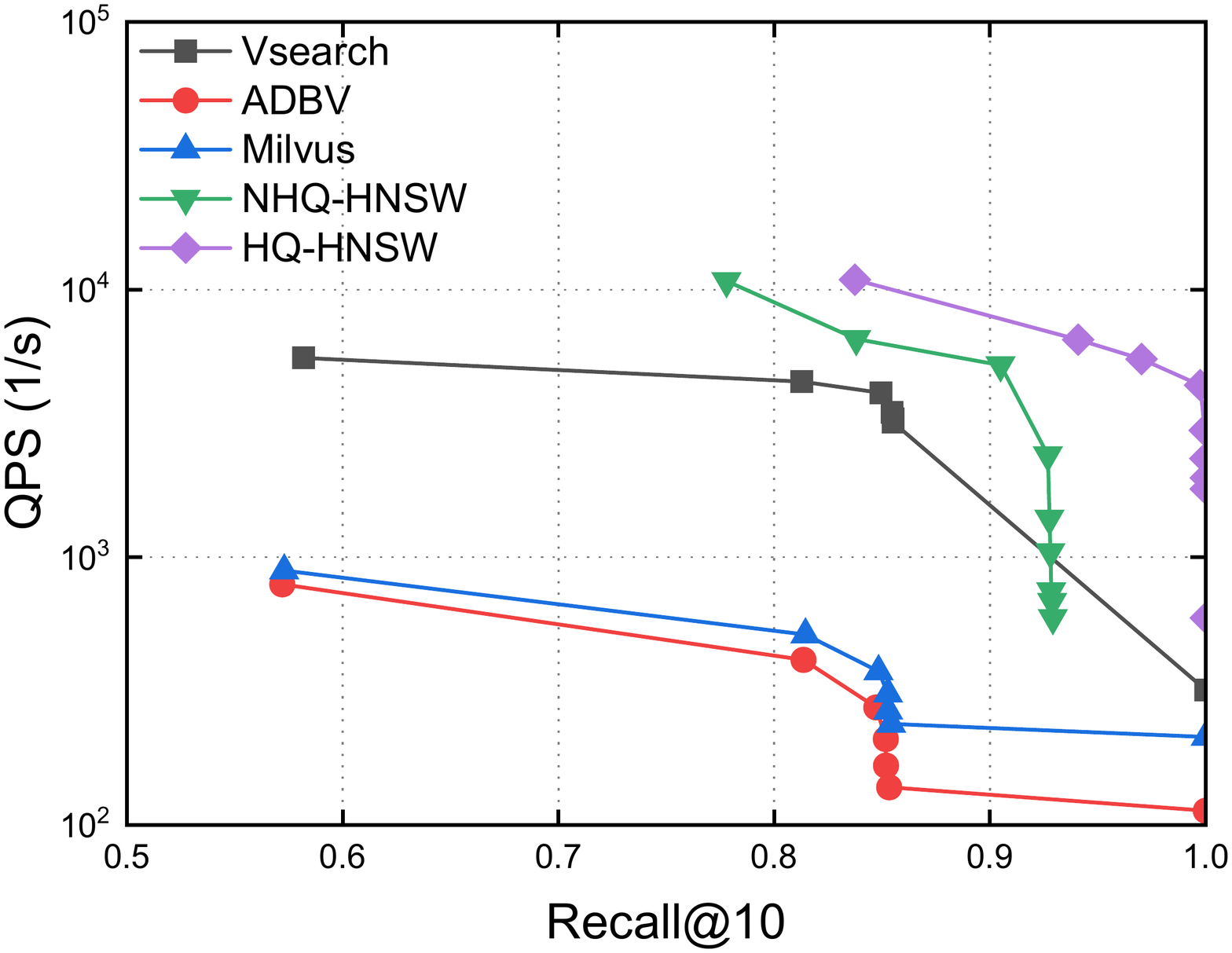}
}
\subfloat[Recall@10 (DEEP-1B)]{
\includegraphics[width=4.2cm,height=2.8cm]{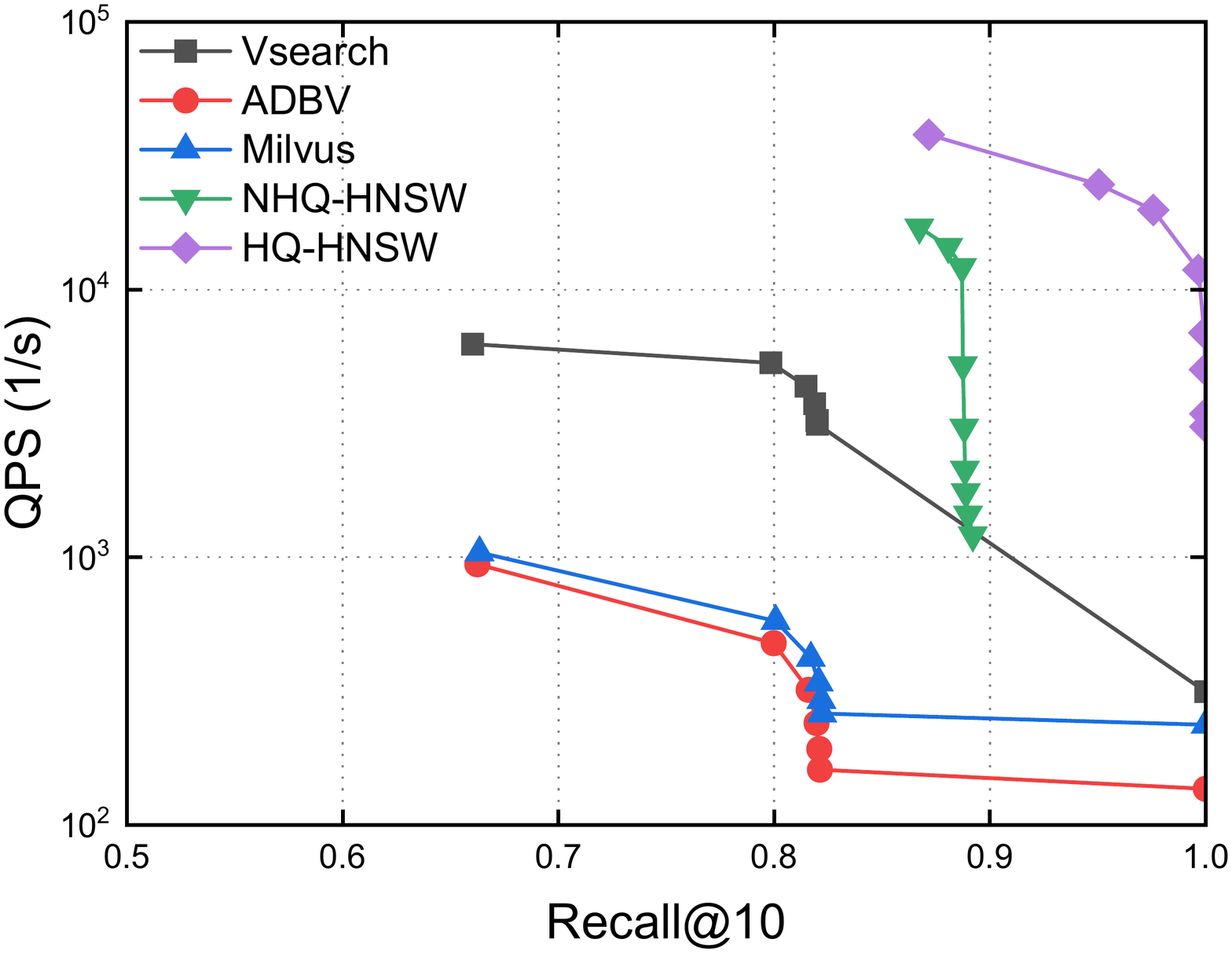}
}
\caption{Speed-recall trade-off on different datasets}
\label{f3}
\end{figure*}

\begin{figure}[t]
\centering
\subfloat[Attribute-Recall@10]{
\includegraphics[width=4cm,height=2.8cm]{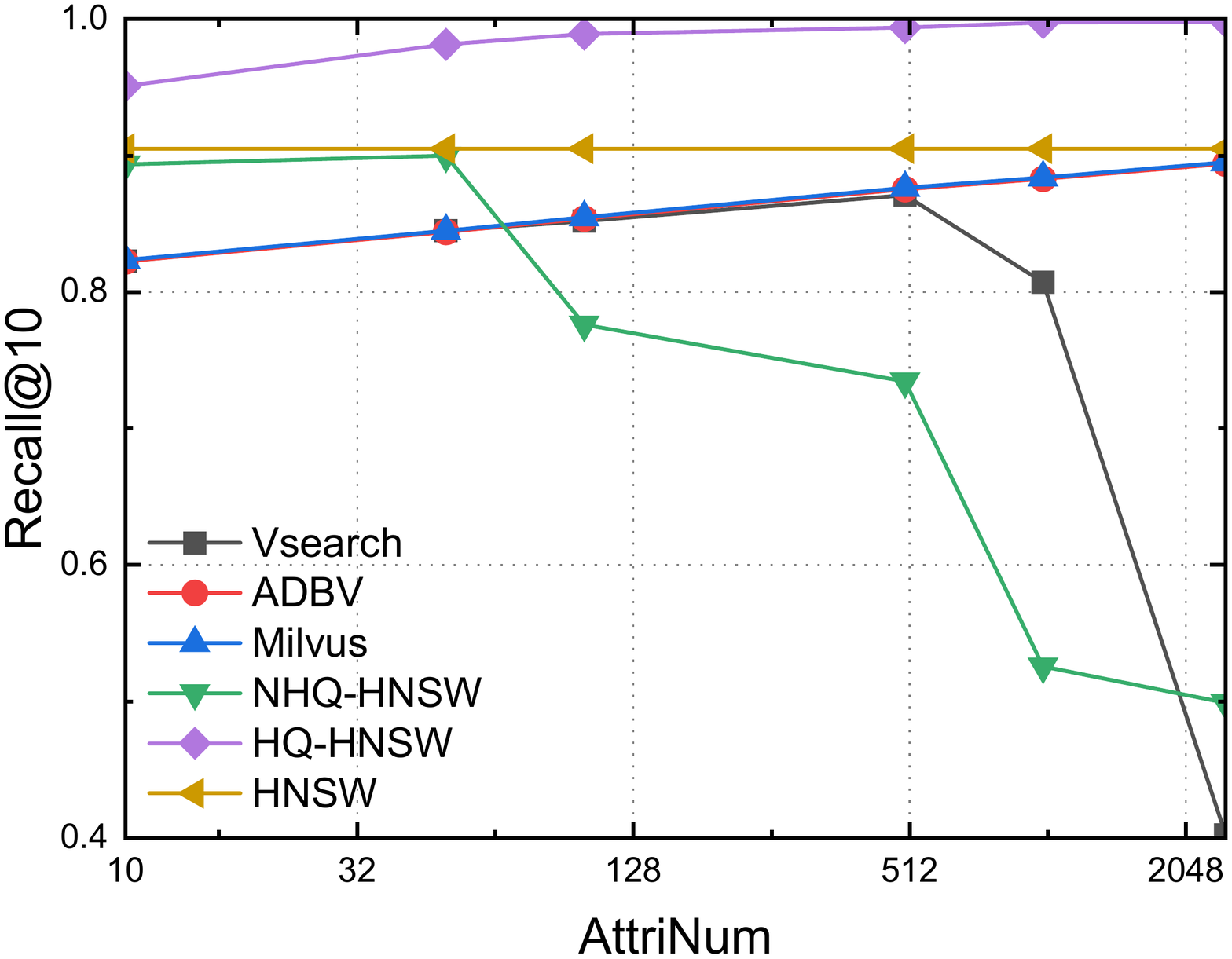}
}
\subfloat[Attribute-QPS(1/s)]{
\includegraphics[width=4cm,height=2.8cm]{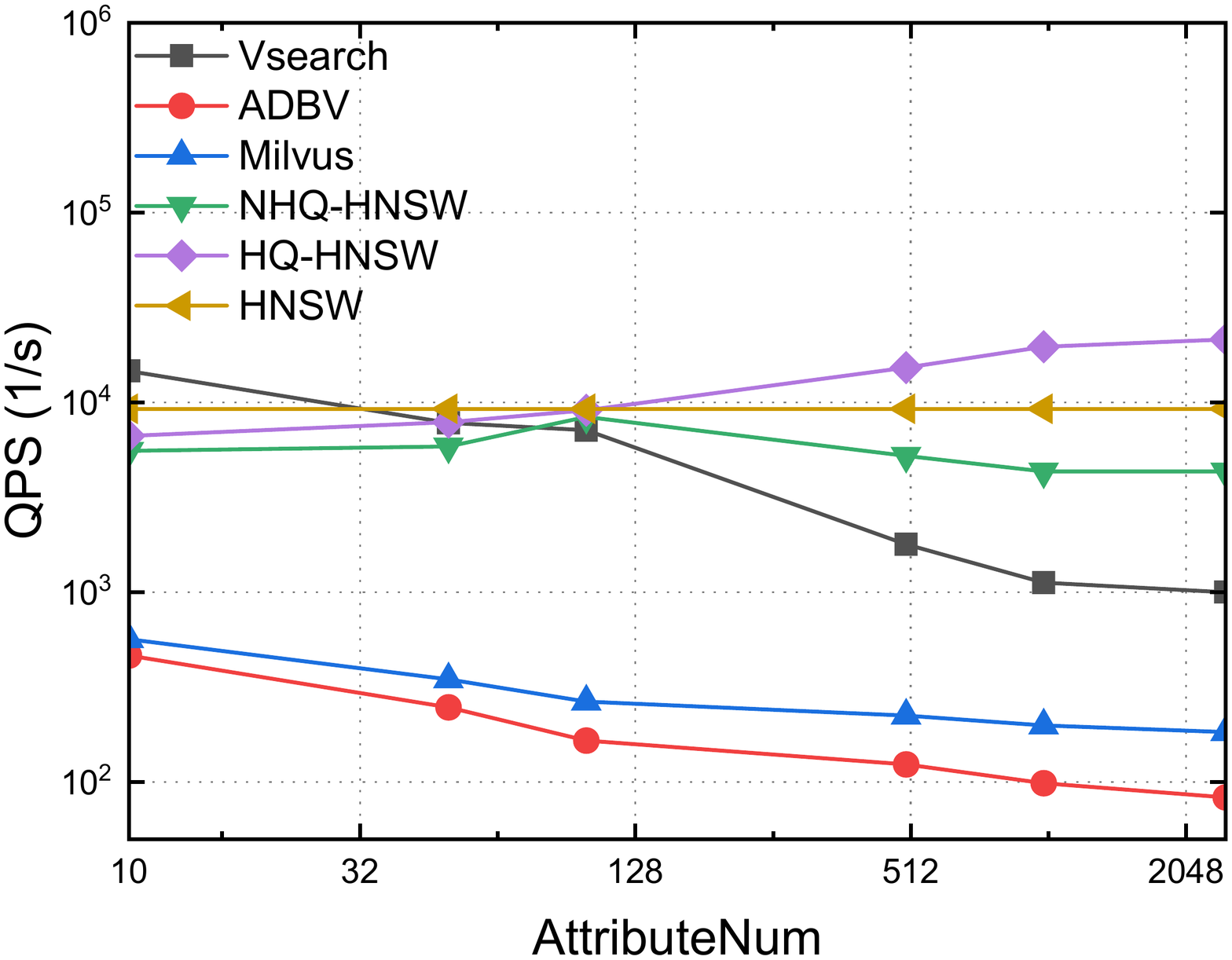}
}
\caption{Recall and Speed performance on GLOVE-1.2M with different scale of attribute constraints}
\label{f4}
\end{figure}

\subsection{Recall-Speed Benchmark}
To evaluate effectiveness of hybrid query processing algorithms in a realistic setting, it is important to performance end-to-end retrieval experiments and compare speed-recall curves. We choose three state-of-the-art two-stage hybrid query processing frameworks(Vearch~\cite{vearch}, ADBV~\cite{adbv}, milvus~\cite{milvus}) and the recently proposed proximity graph-based algorithm (NHQ~\cite{nhq}) as our comparison targets. Our implementations of two-stage frameworks build on product quantization with SIMD based ADC~\cite{MIPADC} for distance computation, which offers a strong baseline to confirm superiority of HQANN. We fix the bit-rate at $dimension \times 4bits$ and carefully tune the search space to observe reasonable recall-speed results. Both HQANN and NHQ are implemented on HNSW~\cite{HNSW}], with the same hyper parameters for all datasets.($M=64, efConstruction=512$). The number of possible attribute constraints we used here is 100.

The performance results on four public datasets are shown in Figure ~\ref{f3}. It demonstrate that HQANN significantly outperforms competing methods in terms of speed and recall rate (up and to the right is better) on all four datasets and it can obtain a recall of 99\% for all datasets. Specially, HQANN is more than ten times faster than other methods to reach the good recall quality (95\%).

\subsection{Robustness Verification}
Then we conduct an experiment on GLOVE-1.2M to compare the robustness with existing hybrid ANNS solutions. For Vearch, we return 100 times more candidates (1000 for Recall@10) for further attribute filtering. For ADBV and milvus, we traverse all datapoints outside the blacklist with product quantization based ADC. For NHQ and HQANN, we fix the same hyper parameters for HNSW($M=64, efConstruction=512, efSearch=80$). In addition, to evaluate the hypothesis that using attributes to limit search space can improve search performance, we attach the search performance of HNSW on GLOVE-1.2M without attribute constraints as a baseline with exactly the same settings. We gradually increase the number of attribute constraints from 10 to 2500, detailed results are shown in Figure ~\ref{f4}. We see that Vearch and NHQ suffer from dramatic accuracy drop, and all other methods become several times slower with the increasing categories of attributes. On the contrary, HQANN can maintain very high recall quality (over 95\% Recall@10) and its efficiency increases as attribute constraints become more complex. It also demonstrates that vector similarity search can benefit from attribute constraints in both latency and recall for its filtering feature. For example, with 2500 attribute constraints, the composite graph is two times faster with 10\% recall gains.

\subsection{Parameter Sensitivity}
Finally, We select some typical scenarios to prove that we do not need to tune these hyper parameters of HQANN for specific datasets. Since $bias$ just need to satisfy $bias \gg w + 3.32$, we fix $bias = 4.32$ and verify the proper ratio of attribute distances to feature vector distances by simply shrinking $w$. Table 1 shows that for simple scenarios such as GLOVE-1.2M with hundreds of attributes, $w=1$ can reach a reasonable recall quality, while for big data scenarios with millions of attribute constrains, shrinking $w$ from 1 to 0.25 brings huge recall benefits. It also proves that further shrinking $w$ cause no recall gains, which means $w = 0.25$ is a suitable critical point for aligning the ratio of two distance parts. We stop reduce $w$ any further in case feature vector part is too small to be completely ignored.

\section{conclusions}
In this paper, we propose HQANN as a new hybrid query processing framework. The new framework leverages navigation sense among attribute constraints, leading to a fusion distance metric which can be flexibly embedded into proximity graph-based ANNS algorithms. Our experiments show superior performance on retrieval recall and efficiency compared with existing hybrid query solutions.

\begin{table}[t]
\caption{Recall@10 comparison with different settings of $w$ on GLOVE-1.2M and merchandise datasets, values in parentheses represent the number of attribute constraints.}
\resizebox{0.47\textwidth}{10mm}{
\begin{tabular}{ccccc}
\hline
\multirow{2}{*}{Dataset} & \multicolumn{4}{c}{Scale Factor $w$}        \\ \cline{2-5} 
                         & $w$=1.00 & $w$=0.50 & $w$=0.25         & $w$=0.10 \\ \hline
GlOVE-1.2M (10)          & 0.986  & 0.987  & \textbf{0.989} & 0.988  \\ \hline
GLOVE-1.2M (100)          & 0.947  & 0.949  & \textbf{0.950} & 0.950  \\ \hline
merchandise-0.2B (0.8M)  & 0.727  & 0.955  & \textbf{0.999} & 0.999  \\ \hline
\end{tabular}}
\end{table}

\clearpage

\bibliographystyle{ACM-Reference-Format}
\bibliography{cikm}

\end{document}